\documentclass[runningheads]{llncs}
\usepackage{graphicx}
\usepackage{todonotes}
\usepackage{amsfonts} 
\begin{document}
\title{Interactive segmentation via deep learning and B-spline explicit active surfaces}
\titlerunning{Interactive deep learning and B-spline explicit active surfaces}
%\authorrunning{H. Williams et al. }
%anonymise for final submission
%\author{H. Williams et al.}
\author{Helena Williams\inst{1,3,4}  \and
Jo\~{a}o Pedrosa\inst{2} \and
Laura Cattani\inst{1}  \and Susanne Housmans\inst{1} \and
Tom Vercauteren\inst{3} \and Jan Deprest\inst{1} \and Jan D'hooge\inst{4} }
\authorrunning{H Williams et al.}

%index{Williams,Helena}
%index{Pedrosa, Jo\~{a}o}
%index{Cattani, Laura}
%index{Housmans, Susanne}
%index{Vercauteren, Tom}
%index{Deprest, Jan}
%index{D'hooge, Jan}

%\institute{**** \and *****\and ***** \and *****}
\institute{ Department of Obstetrics and Gynaecology, University Hospitals Leuven, Belgium  \and INESC TEC - Institute for Systems and Computer Engineering, Technology and Science, Porto Portugal \and
School of Biomedical Engineering \& Imaging Sciences, King’s College London, UK
\and
Department of Cardiovascular Sciences, KU Leuven, Belgium
}

%\author{Helena Williams \and
%Jo\~{a}o Pedrosa  \and
%Laura Cattani \and Susanne Housmans \and Tom Vercauteren \and Jan Deprest \and Jan D'hooge}
%
\authorrunning{H. Williams et al.}
%\institute{********; ********; *********}
%\institute{KU Leuven, Belgium; King's College London, United Kingdom; INESC TEC - Institute for Systems and Computer Engineering, Technology and Science, Porto Portugal}
%
\maketitle            % typeset the header of the contribution
\begin{abstract}
Automatic medical image segmentation via convolutional neural networks (CNNs) has shown promising results.
%in literature and clinical applications.
However, they may not always be robust enough for clinical use. Sub-optimal segmentation would require clinician's to manually delineate the target object, causing frustration. To address this problem, a novel interactive CNN-based segmentation framework is proposed in this work. The aim is to represent the CNN segmentation contour as B-splines by utilising B-spline explicit active surfaces (BEAS). The interactive element of the framework allows the user to precisely edit the contour in real-time, and by utilising BEAS it ensures the final contour is smooth and anatomically plausible. This framework was applied to the task of 2D segmentation of the levator hiatus from 2D ultrasound (US) images, and compared to the current clinical tools used in pelvic floor disorder clinic (4DView,
%Point and 4DView Trace,
GE Healthcare; Zipf, Austria). Experimental results show that: 1) the proposed framework is more robust than current state-of-the-art CNNs; 2) the perceived workload calculated via the NASA-TLX index was reduced more than half for the proposed approach in comparison to current clinical tools; and 3) the proposed tool requires at least 13 seconds less user time than the clinical tools, which was \textit{significant} (p=0.001).     
%\keywords{semantic segmentation  \and interactive segmentation \and B-spline explicit active surfaces.}
\end{abstract}
\section{Introduction}
Medical image segmentation of anatomical structures can be used for disease diagnosis \cite{levhiunet}. Manual segmentation requires
expertise, time and is prone to error, therefore, automatic methods are desirable. Deep learning-based solutions with convolutional neural networks (CNNs) have been extensively explored \cite{Deeplearningreview,doi:10.1146/annurev.bioeng.2.1.315}. However, their impressive average performance has not yet led to wide clinical adoption \cite{fda}.
% does not coincide with direct clinical implementation \cite{fda}.
%A list of AI-equipped medical devices approved by the USA FDA up until July 2020 can be found on ...[ref]. According to this paper 60 products have been approved %In September 2020, only 64 Artificial Intelligence (AI) or Machine Learning (ML) based medical devices or algorithms had been FDA approved, some of which use CNN segmentation to locate structures from medical images \cite{fda}. 
%It can be assumed in order for segmentation algorithms to be implemented they must be evaluated on a large, clinical dataset from multiple-centers and perform with high accuracy. 
%This may be due to the nature of automatic segmentation, as they do not achieve consistent highly accurate and robust results, to be clinically useful and reliable \cite{automedimageseg}. 
Medical images pose serious challenges to automatic methods, as they can be sensitive to small differences between training and testing data, due to such factors as image quality, imaging protocols (i.e. imaging acquisition discrepancies), pathology, and patient variation \cite{https://doi.org/10.1002/uog.11190,patientvar,DBLP:journals/corr/abs-1710-04043,DBLP:journals/corr/abs-1903-08205}. Therefore, it is important for clinical impact and acceptance, to be able to recover from a poor result and address the limitations of automatic segmentation. As the clinician remains liable for the measurements obtained for diagnosis, if the automatic method is incorrect, it is the responsibility of the clinician to identify the problem and correct the segmentation. 
%Automatic methods that are not easy to interact with, or not possible to interact with can make it frustrating for the user, which can reduce trust between the clinician and the algorithm.
Interactive segmentation with an intuitive mechanism, for smart correction of poor segmentation, may solve these problems, and give liability to the clinician without them having to manually re-segment, which is not time efficient and may cause frustration. 
This work is motivated  to combine state-of-the-art CNN segmentation with an user interaction tool, which allows the clinician to view, correct (if needed) and save the desired segmentation.

An extensive range of CNN-based interactive methods have been proposed~\cite{interactivesegsurvey}, 
%such as
exploiting
bounding boxes \cite{DBLP:journals/corr/RajchlLOKPBKR16}, scribbles \cite{DBLP:journals/corr/LinDJHS16,wangUGIR}, extreme points \cite{DBLP:journals/corr/abs-1711-09081} or clicks \cite{jang2019interactive}. These achieved higher accuracy and robustness than their automatic counterparts, however, they can require a high cognitive load and understanding. 
In addition, the user still relies on the CNN to segment correctly and is not always able to edit the contour precisely, in an adequate and time efficient manner. 

In this paper, an interactive segmentation tool for 2D semantic medical image segmentation is proposed. The tool is composed of three stages. In the first stage, a CNN automatically obtains an initial segmentation, this feeds as an initialisation to an active contour segmentation framework called B-spline explicit active surfaces (BEAS) \cite{beasbarboa2012}, which smooths the contour to be more biologically plausible (acting as a post-processing step), and thirdly a novel algorithm which allows the user to interact with the contour in real-time was implemented. %on a graphical user interface (GUI) called Beyond, the GUI loads the image and segmentation to allow for real-time changes, based on the clinician dragging points of the contour to the desired position. %The proposed tool benefits from advances in CNN's, active shape modelling through B-spline explicit active surfaces (BEAS) [beas] and intuitive, real-time, interaction [daniel b interactive beas]. 

%%Compared with existing interactive segmentation methods, the proposed pipeline has several appealing properties.
%First, it uses state-of-the-art CNN models to provide an initial contour automatically. 
%BEAS has been used in the 2D segmentation of the levator hiatus, however, the method was semi-automatic and required several user defined points [ref nikhil]. 
%%Firstly, the CNN could be replaced by any deep or machine learning automatic or semi-automatic 2D segmentation method. 
%The tool can also be used to edit manual segmentations prior to CNN training, in order to have a biologically accurate contour, or to quickly adjust contours from less trained clinicians. 
%%Secondly, BEAS can smooth the CNN contour to be more biologically plausible.
%, as CNN segmentation is pixel-wise occasionally there may be spikes, pixels missing or anatomically incorrect features. 
%%The BEAS framework limits the topology, this may be seen as introducing a shape constraint which is not easily feasible in CNN segmentation. Thirdly, the proposed algorithm allows for real-time interaction and therefore, a short user time. In addition, the interactive element does not rely on deep learning and GPU power, thus it can be run on any machine. 
The proposed approach is compared with manual tools used in clinic
for 2D segmentation of the levator hiatus in pelvic floor disorder assessment:
``Point" and ``Trace" both available on the ultrasound (US) software 4DView (GE Healthcare; Zipf, Austria), and compared with a state-of-the-art scribble-based approach referred to as UGIR \cite{wangUGIR}.
The segmentation methods are evaluated on 30 2D US images, the time taken to segment to a clinically acceptable level, and the perceived workload are measured and compared. 
%Previous literature shows several automatic and semi-automatic methods that accomplish high quality segmentations, however, they lack interaction and have not been implemented into clinic. Nikhil Sindhwani utilised BEAS and user-defined clicks to segment the levator hiatus, this semi-automatic approach was more time-consuming than CNN based approaches and  failed to accurately segment the hiatus in several cases [ref]. Several papers demonstrate CNN based automatic segmentations of the levator hiatus from 2D US images, however, they did not include a tool to edit and adapt poor segmentations, thus it would leave clinicians to manually segment if incorrect. By combining both BEAS and CNNs we believe to improve the semi-automatic approach by making it fully automatic, and improve CNN based methods by making it interactive and allowing the user to adapt the contour if necessary.  
The contributions of this work are four-fold:
1) A novel CNN-based interactive framework for 2D segmentation is proposed; 2) the interactive element works in real time and requires less user time and perceived workload than clinical methods and UGIR; 3) the method utilises the BEAS framework to ensure the final contour is more biologically plausible than the CNN segmentation, acting as a novel post-processing method; and 4) a new energy term is introduced that is dependent on the probability map of the CNN output, which has not been utilised in BEAS before. 
 %the pipeline is evaluated on 30 US images by two expert clinicians and compared to the current manual tools available in clinic, we demonstrate that the tool is faster, intuitive and provides clinically acceptable contours.  
\section{Material and Methods}
\subsection{Proposed pipeline}
\begin{figure}[t!]
\centering
\includegraphics[width=.9\textwidth]{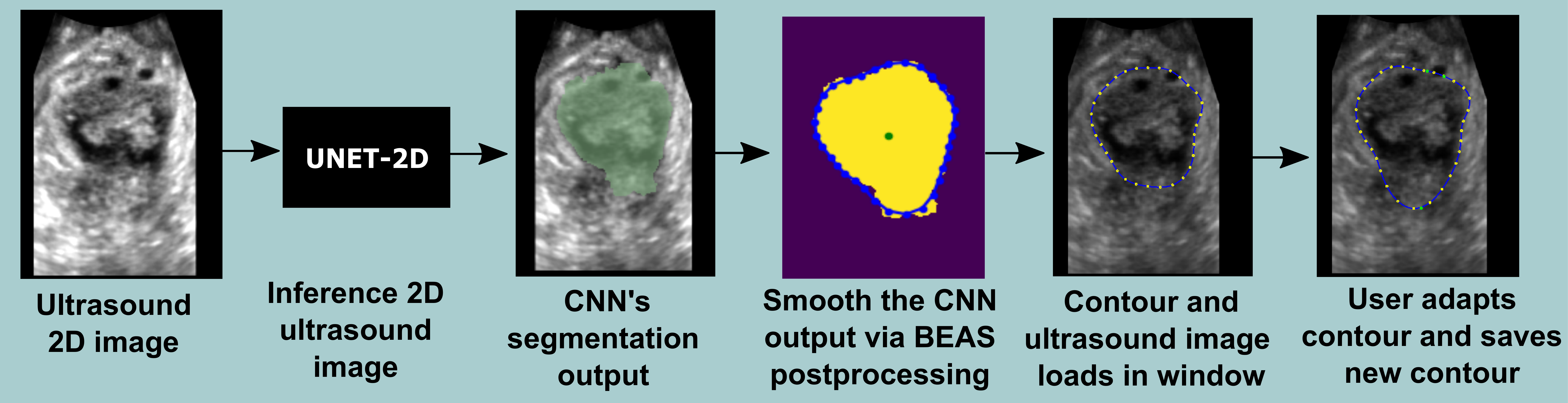}
\caption{The proposed pipeline that accepts a 2D US, segments the target object via CNN segmentation, smooths the segmentation and loads it in a window, to allow the user to adapt the contour via BEAS interaction.} \label{pipeline}
\end{figure}

The proposed pipeline is composed of three sequential parts: a 2D CNN which segments the target object (levator hiatus) from the US image; a BEAS-based post-processing method which smooths the CNN segmentation and represents the segmentation boundary as a B-spline explicit active surface; and a novel algorithm (implemented in a graphical user interface (GUI) referred to as Beyond), that allows the user to adapt the contour in real-time while benefiting from BEAS's active 
%shape
model properties. The framework is shown in Fig.  \ref{pipeline}.
The first task of the pipeline, automatically defined the levator hiatus from the US image. This elaborates from previous work where 2D U-Net was used \cite{levhiunet,liunetlevhi}. The segmentation is fed as input to the following task. 
\subsubsection{BEAS-based smoothing}
The second task utilises the BEAS framework \cite{beasbarboa2012}. A 2D version of BEAS is applied to the CNN segmentation after thresholding, to represent the CNN segmentation boundary as B-splines. The concept of BEAS, is to regard the boundary of a target object (i.e. the CNN segmentation) as an explicit function, where one of the coordinates of the boundary, is given explicitly as a function of the remaining coordinates. % i.e. $x_1=\psi(x_2,...,x_n)= \psi(x^*)$.  
 As the contour is a closed 2D object, the boundary can be represented in the polar domain, and the contour radius is represented as an explicit function of the polar angle, $i.e.  \rho = \psi(\theta)$. Inspired by Bernard et al. \cite{4895341}, the explicit function $\psi$ can be expressed as a linear combination of B-spline basis functions  \cite{SPLINES,4895341,beasbarboa2012},
\begin{equation}
    %x_1 = \psi(x,...,x_n) =
    \rho = \sum _{\textbf{k}\in \mathbb Z^{n-1}} c[\textbf{k}]\beta^d \left( \frac{\theta}{h} - \textbf{k} \right).
\end{equation}
$\beta^d (.)$ is the uniform symmetric n-1-dimensional B-spline of degree d. $\psi(\theta)$ is separable and built as the product of n-1 1D B-splines. The knots of the B-splines are located on a regular grid defined on the polar coordinate system, with regular spacing given by $h$.  The B-spline coefficients are gathered in $c[\textbf{k}]$.

BEAS assumes that all coordinates of the boundary are visible from a fixed origin, which is a good approximation for this structure that tends to have a pear-like convex shape. Before refinement of the BEAS contour,
%can begin,
the initial circular contour must be defined by parameters, such as the fixed origin and an initial radius. In this work, they are based on properties of the CNN segmentation output. The origin is defined as the center of mass of the CNN output, and the initial radius, $r_c$, is the average radius of the CNN segmentation output. 

The initial contour can then refine and evolve towards the boundary of the CNN segmentation through the minimisation of a segmentation energy functional. To achieve this, a general localised region-based energy functional for level-set segmentation \cite{localyezzi} was used.  Barbosa et al. \cite{beasbarboa2012} adapted these localisation strategies for BEAS in terms of B-spline coefficients, and the expression of the energy gradient is given as,
\begin{equation}
   \nabla_c E= \frac{\partial E}{\partial c[k]} = \int_\Gamma g(\theta)\beta^d \left(\frac{\theta - hk}{h}\right) d\theta.  
\end{equation}
 The function $g(\theta)$ represents the features of the object to be segmented and is evaluated over the boundary $\Gamma$. 
 %The computation of the energy gradient wrt. B-spline coefficients may be interpreted as convolving  $g(x^*)$ with B-spline and sampling the result with period h. 
In this work, the energy function used was the Localised Yezzi Energy, proposed by Lankton and Tannenbaum et al. \cite{localyezzi,Yezzi2002AFG}. This energy depends on the average intensity of the CNN output inside and outside the evolving B-spline contour. The contour evolves to have the maximum separation between them. For Localised Yezzi the feature function is given as, \begin{equation}
     g(\theta)= \frac{\big(I(\psi(\theta),\theta)-u_\theta\big)^2}{A_u}- \frac{\big(I(\psi(\theta),\theta)-v_\theta\big)^2}{A_v}.
 \end{equation}
$A_u$ and $A_v$ represent the areas inside and outside of the contour, respectively; and $u_\theta$ and $v_\theta$ are the mean intensities inside and outside the evolving contour at the polar angle, $\theta$, respectively. $I(\psi(\theta),\theta)$ corresponds to the image value (i.e. CNN output) at position $(\rho,\theta$). 
The Yezzi energy relies on the assumption that the interior and the exterior of the contour have the largest difference in average intensities. This is a good assumption for this work, as the goal is to represent the CNN segmentation output as a smooth B-spline contour. The final B-spline coefficients are saved and used in the following section. 
\subsubsection{Interaction framework}
Finally, the contour formed from the previous step and the corresponding US image are loaded in a window, where the user can interact with the contour. %A screenshot of the Beyond GUI can be shown in fig.~\ref{fig:guibeyond}.
% \begin{figure}[hbt!]
%\centering
%\includegraphics[width=1\textwidth]{gui beyond.pdf}
%\caption{An annotated screenshot of Beyond's GUI, the contour is shown as a blue line, the yellow dots show the knots that define the B-spline contour, and the green dots show the user-defined points.} \label{fig:guibeyond}
%\end{figure}
%This is driven by BEAS, however, the energy function is not solely dependent on one energy term. 
In the interactive framework, the energy function driving BEAS, is compounded of three energy terms:  Localised Yezzi of the US image, $E_U$,  Localised Yezzi of the CNN probability map, $E_{CNN}$ and an interactive energy function, $E_{i}$. The total energy is given as: 
\begin{equation}
    E_{total}= \alpha E_{U} + \beta E_{CNN} + \gamma E_{i},
\end{equation}
where $\alpha$, $\beta$ and $\gamma$ are hyper-parameters. Here, the initialised contour is determined by the evolved B-spline coefficients from the previous section, and it will evolve with each user-interaction to minimise the energy function defined. 
Finally $E_i$, is based on a 2D version of reported work \cite{interactivebeas}, where user-defined coordinates interact with the B-splines. The user can create markers where they want the contour to pass through, these act as anchors attracting the contour.  %where $\psi(x^*)$ corresponds to a contour whose radius is a function of the azimuthal angle, i.e. $\rho=\psi(\theta)$. 
A point introduced by the user can be expressed as $p_{user}=(\rho_{user},\theta_{user})$. The energy function penalises the parametric distance, D, between the current boundary position and $p_{user}$ at each B-spline knot. Therefore, D is defined as $ D=(\rho-\rho_{user})^2$.
%The contour can evolve towards the user's desired position. 
The energy term driving the contour towards the user-defined points was proposed in 3D by Barbosa et al. \cite{interactivebeas}, where its minimisation with respect to B-spline coefficients, $c[\textbf{k}]$ was demonstrated. $E_{user}$ is defined as:
\begin{equation}
 E_{user}= \int_\Gamma \delta (\theta- \theta_{user})(\psi(\theta)-\rho_{user})^2 d \theta,
\end{equation}
where $\delta(\theta-\theta_{user})$ corresponds to the Dirac function which is non-zero only at the position $\theta = 
\theta_{user}$. 
When multiple user-defined points are present, the sum of the parametric distance between the current contour, $\Gamma$ and the user-defined points is used. This evolves the contour towards multiple user points, detailed information can be found in the paper by Barbosa et al. \cite{interactivebeas}.
As the computational load is small, there is real-time feedback of the effect the modifications make. 
%$N_p$ defines the number of points introduced by the user, $\partial (\theta- \theta_{user})$ corresponds to the 2D Dirac delta function which is non-zero at the position $x^*=(\theta_{user})$

%Beyond is based on a 2D version of reported work [barbosa interactive], where user defined coordinates interacted with the B-splines. In beyond the user will drag the contour towards the desired position leaving a green coordinate marker. These markers act as anchors attracting the boundary. This allows the user to interact with the contour and, as BEAS computational load is small, there is real-time feedback of the effect the modifications make. The points introduced by the user are expressed in polar coordinates.   

%while BEAS is continuously running and further refining the contour for a given time. 

%In this section the energy function has changed from Yezzi to one made specifically for this work. Previous work has developed an energy function that combines Yezzi coupled with interactive   

%The B-spline contour may change slightly when loaded as the energy function has changed from Yezzi to one made specifically for this work. 

%In beyond one can interact with the B-spline contour by dragging one of the B-splines to the desired position.  
\subsection{Data collection }
Analysis of anonymised, archived US images was retrospective, so no ethics committee approval was required by the institute.
%KU Leuven, Belgium. 
The CNN was trained on a dataset of 444 2D US images from 213 patients, and corresponding ground truth labels of the levator hiatus. The training dataset comprised of two sets of archived clinical images with expert annotations, acquired by several operators. One dataset used for training, was a private dataset supplied by (GE Healthcare; Zipf, Austria) and the second dataset was a private dataset used in previous studies. %\cite{levhiunet,SindhwaniNikhil2016SOoL}. 
%This allows the CNN to learn a variety of acquisition parameters, image qualities and segmentation styles. 
%Within the training dataset a subset of 91 2D US images with annotation were used in our previous studies (Bonmati, et al. 2018, Sindhwani, et al. 2016). A subset of 354 2D US images with annotation were from a private database from GE Healthcare (zipf, Austria), and finally 55 2D US images and annotations were from the pelvic floor disorder clinic at UZ Leuven, Belgium. 
400 images were used for training and 44 were used for validation. 
The test data included a randomised selection of 30 anonymised 2D US images from 10 symptomatic women assessed at the pelvic floor clinic between March and May, 2019 at *****.
%UZ Leuven, Belgium.  
The US images were obtained from Transperineal US volumes acquired following the clinical protocol defined by Dietz et al. \cite{https://doi.org/10.1002/uog.1899} on the Voluson E10 US system (GE Healthcare; Zipf, Austria). The 2D planes that were used to segment and assess the levator hiatus were manually determined by an expert clinician, at rest, during the Valsalva manoeuvre and contraction. 
%The Valsalva manoeuvre is achieved by forcefully exhaling against a closed airway, and the condition called ballooning hiatus (where the area is larger than $25cm^2$ \cite{Avulsioninjury}) can be determined in this movement, this can indicate if the patient has pelvic organ prolapse.  Supplementary information of the patient demographic  used for evaluation can be found in the appendix.  
\subsection{Experimental details}
Two clinical experts with over 4 years experience in pelvic floor US, participated in the experiment. They segmented the levator hiatus on 30 2D US images using 4DView Trace (GE Healthcare; Zipf, Austria) , 4DView Point (GE Healthcare; Zipf, Austria) and the proposed tool. Prior to the experiment the experts were given a tutorial how to use the new tool. 4DView Trace and Point can be found on the `measure - generic area' function of 4DView (GE Healthcare; Zipf, Austria). %\subsubsection{Manual segmentation tools}
 In 4DView Trace (GE Healthcare; Zipf, Austria), the contour starts once the user clicks the US image, and it will follow the user's cursor around the levator hiatus until the user clicks on the US a second time. In 4DView Point (GE Healthcare; Zipf, Austria) the tool will trace the hiatus by the user defining multiple points around the levator hiatus with mouse clicks. The lines that connect the points are straight, therefore, the output segmentation is generalised and not anatomically accurate (i.e. sharp lines). 4DView Trace and 4DView Point (GE Healthcare; Zipf, Austria) may be referred to as Trace and Point respectively in this paper. Uncertainty-Guided Efficient Interactive Refinement (UGIR)  utilises an interaction-based level set for fast refinement of segmentations \cite{wangUGIR}, based on scribbles. The same CNN was used as the proposed model and scribbles were created in 3D Slicer \cite{3dslicer,FEDOROV20121323}.  
%\subsubsection{Evaluation metrics}

The main aim was to compare the perceived subjective workload of the clinical tools and UGIR against the proposed tool. Therefore, half way through the experiment (after 15 segmentations) and at the end of the experiment, the perceived workload was subjectively evaluated by each expert and for each segmentation technique. To do this the National Aeronautics and Space Administration Task Load Index (NASA-TLX) was used \cite{doi:10.1177/154193120605000909}.
%%NASA-TLX \cite{doi:10.1177/154193120605000909} is the most extensively utilised subjective questionnaire for mental workload assessment.  The NASA-TLX includes six subjective sub-scales: mental, physical, and temporal demands, effort, perceived performance and frustration. Descriptions for each sub-scale can be found in the supplementary information. The expert was asked to rate their score for each sub-scale on an interval scale from low (1) to high (20). NASA-TLX also employs a paired comparison procedure, which involves presenting 15 pairwise combinations to the participants, and asking them to elect the sub-scale from each pair that has the most effect on the workload during the task under analysis. 
%This tool accounts for two potential sources of variability: the difference in workload definition between experts and the difference in sources or workload between the tasks.
%A final perceived weighted mental workload score is obtained which ranges from 0 (low) to 100 (high). 
Finally, the time taken for the expert to segment/edit the levator hiatus contour to a clinically accepted level was measured for each segmentation and compared. 
%\subsubsection{Statistical analysis}
%To evaluate whether the time taken was statitically significant, a paired t-test was used to test several null hypotheses. The first was that the newly proposed tool 
\subsection{Implementation details}
The proposed tool was implemented on a Windows desktop with a 24GB NVIDIA Quadro P6000 $($NVIDIA, California, United States$)$. %\subsubsection{2D CNN segmentation}
The CNN was implemented using NiftyNet \cite{DBLP:journals/corr/abs-1709-03485}, training and inference were ran on the GPU. The network architecture was an adaptation of 2D U-Net \cite{DBLP:journals/corr/RonnebergerFB15} with half the number of features $[32,64,128,256,512]$. An Ada{m} optimiser, ReLU activation function, weighted decay factor of $10^{-5}$ and batch size of 64 were used. Whitening and histogram normalisation (i.e. when the image was set to have zero-mean and unit variance) were applied to reduce the effects of noise \cite{7808140}. A Dice loss function was used, with a learning rate of $10^{-5}$. The data augmentation used were: elastic deformation (deformation sigma = 5, number of control points =4), random scaling (-20\%,+20\%), vertical `flipping' and an implementation of mixup \cite{DBLP:journals/corr/abs-1710-09412}. Validation of the network training was performed every 250 epochs and the CNN trained for 12,000 epochs. The CNN model from epoch 10,000 was used at inference, as the validation loss function was lowest. The CNN hyper-parameters were determined based on literature \cite{levhiunet} and the performance of the training dataset.  
%\subsubsection{Post-processing BEAS parameters}

BEAS optimisation was ran on the CPU. In task 2, the size of the neighbourhood used to estimate the local intensity of the image was set to 100 pixels (i.e. $\approx 30mm$), allowing the contour to recover from a bad initialisation. For both tasks the BEAS contour was discretised into $32$ points (i.e. knots) along the polar angle direction, causing the scale parameter, h, to be implicitly fixed to 1. The B-spline coefficients, $c(\textbf{k})$, are gathered in a 1D index array, spanning the polar domain with $32$ B-spline coefficients. For interactive BEAS, in (4) $\alpha$ = 0.5, $\beta$= 0.3 and $\gamma$ = 3.  The size of the neighbourhood used to estimate the local intensity was set to 10 pixels ($\approx3mm$). This is low to avoid the contour evolving before user-interaction. Otherwise the contour may evolve towards bright regions of the US in order minimise the energy function. The hyper-parameters used for BEAS were determined by a grid search method where the range was guided by literature \cite{beasbarboa2012} and evaluated by assessing the performance of the training dataset. 
%The mesh that the contour follows is determined by the number of points and the parameter 'scale'. The number of points the BEAS mesh has was set to 32 and must be equal to $x*2^{(scale+1)}$, where x is an integer larger than 0, these represent the number of splines and knots to the contour has, and the scale was set to 1, which represents the level of smoothing to apply. The segmentation parameters depend on 'normal depth' and 'time step'. At each mesh point BEAS reads the image intensity along the normal of the mesh to evaluate the gradient. The length of these normals is controlled by 'normal depth'. A large number means the contour can recover from a bad initialisation, but if it is too large the contour can be pulled towards structures that are not the target object, in thi work normal depth was set to 100, as the only pixels on the image corresponded to the CNN segmentation. Final the time-step which controls how much the surface moves per iteration was set to 1, again a larger number helps the segmentation optimise if the initialisation is not optimal but makes the segmentation less accurate.   
\section{Results}
\begin{figure}[t!]
\begin{minipage}{0.49\textwidth}
\centering
\includegraphics[width=0.8\linewidth,height=3.5cm]{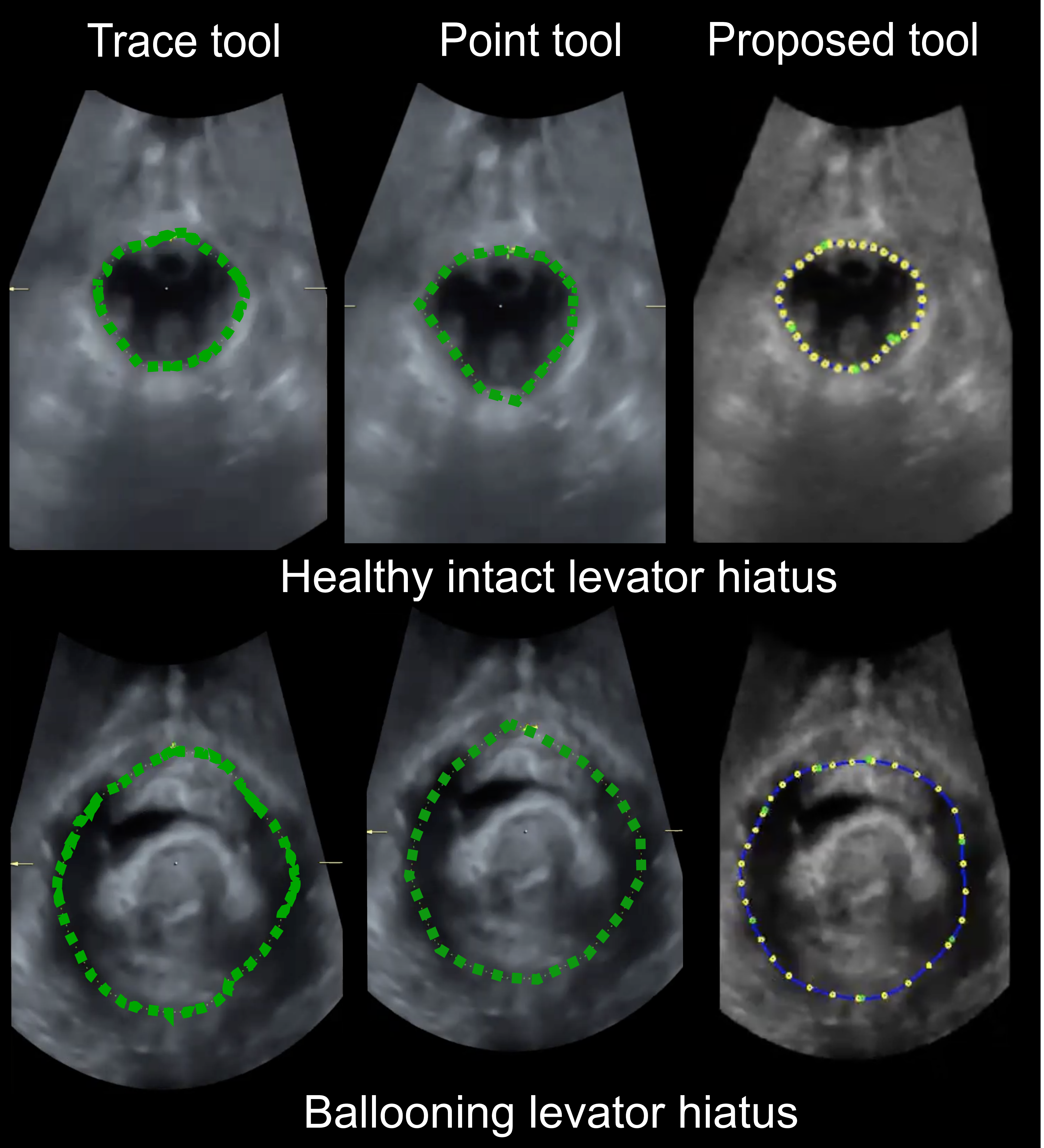}
\caption{Visual representation of the levator hiatus segmentation obtained with the clinical tools and proposed pipeline, in a healthy patient at contraction and a patient with ballooning hiatus at Valsalva.   }
\label{fig:images}
\end{minipage}
~
\begin{minipage}{0.49\textwidth}
\centering
\includegraphics[width=1\linewidth, height=3.8cm]{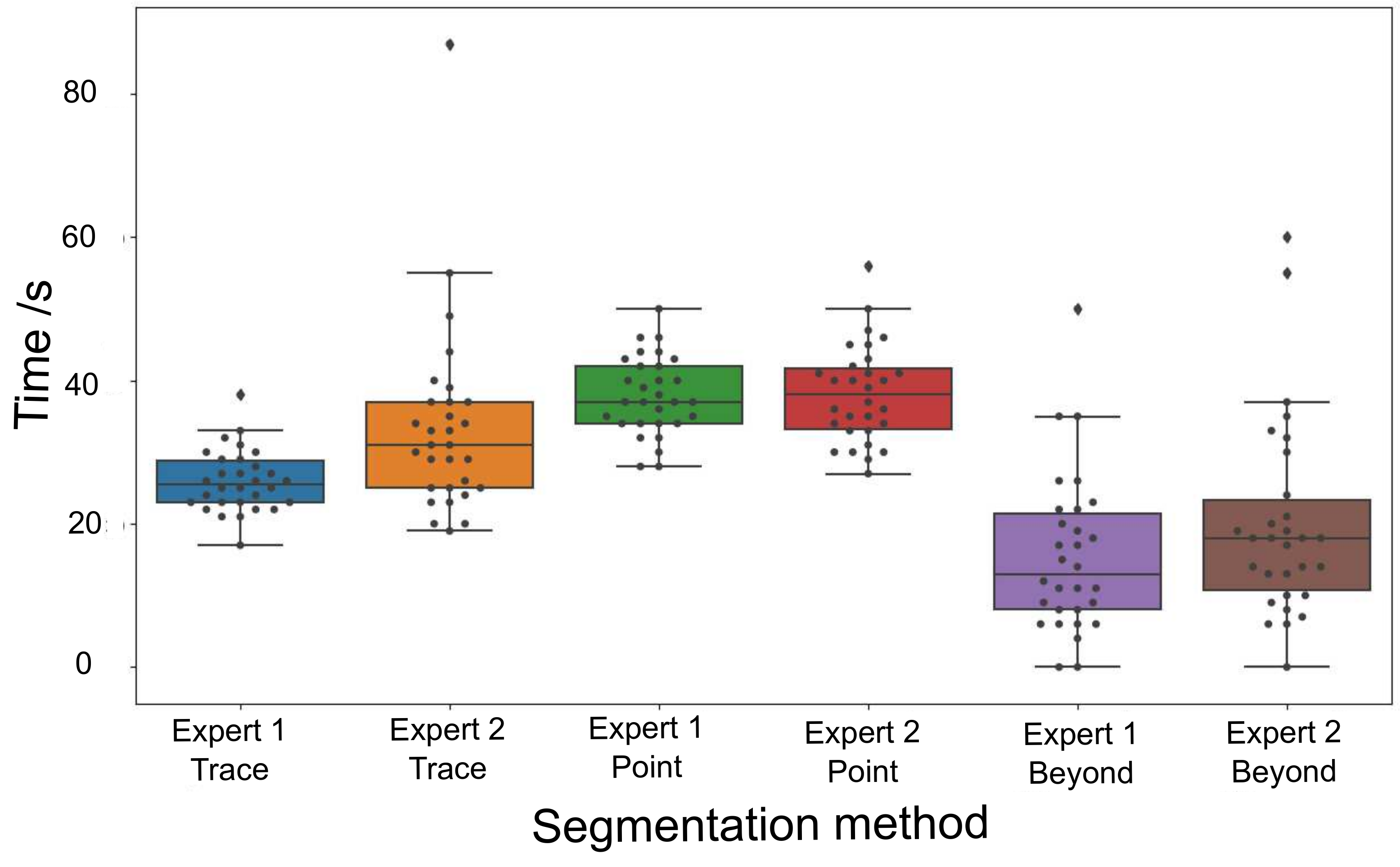}
\\
\caption{Time taken to delineate the levator hiatus to a clinically acceptable level with 4DView Trace, Point (GE Healthcare; Zipf, Austria) and the proposed tool.} \label{time_results}
\end{minipage}
\end{figure}

%The experts with a combined number of 9 years' of experience in pelvic floor disorder assessment completed a training session for using the proposed tool prior to the experiment. The expert then delineated the levator hiatus outline on 30 2D US images from 10 patients.
 Fig. \ref{fig:images} shows examples of the segmentation obtained via the clinical tools and the proposed pipeline. 
%Fig X shows segmentation of the hiatus on a) a  healthy `normal' hiatus, b) a ballooning hiatus and c) a unilateral hiatus, by both experts using all tools. 
The experts agreed that the proposed tool accomplished a clinical acceptable standard for hiatal diagnosis for almost all 2D US images (29 images), thus the proposed tool achieved a `clinical acceptability' of $97\%$. However, only 2 and 1 CNN + BEAS post-processing segmentation's required no editing from expert 1 and 2 respectively, equalling a `clinical acceptability' of $5\%$. The `clinical acceptability' of the CNN alone was $2\%$, and the `clinical acceptability' of UGIR was $27\%$.

Fig. \ref{time_results} shows the time taken for expert 1 and 2 to delineate a segmentation of the levator hiatus to a clinically acceptable standard for diagnosis, using the clinical tools and proposed method. The recorded time of the proposed pipeline does not include CNN inference time, to be solely dependent on user interaction time and independent of the CNN's performance. The average CNN inference time was $4.90\pm1.58$ seconds. The time taken using the UGIR method was $63.77 \pm 31.27$ seconds. 
%The point tool achieved a mean time of 37.90$\pm5.37 $ and 38.20$\pm 6.69$ seconds for expert 1 and 2 respectively, the trace tool achieved a mean time of 25.97 $\pm $4.23 and 33.4 $\pm$ 13.00 seconds for expert 1 and 2 respectively, and finally the proposed tool achieved a mean time of 15.53$\pm$10.96 and 19.93$\pm$ 13.45 seconds for expert 1 and 2 respectively.  
The time taken to edit the levator hiatus contour with the proposed tool was \textit{`significantly lower'} (paired t-test, p$\leq$0.001) than the clinical tools and state of the art method, UGIR. The time taken decreased over the experiment, as the combined mean of the first 15 segmentations was 19.33 $\pm$ 11.93 seconds and for the final 15 segmentations was 16.13 $\pm$ 11.40 seconds. 
Table 1 shows the mean NASA-TLX scores of both experts, and table 2 shows the NASA-TLX scores half way through the experiment (attempt 1), and at the end of the experiment (attempt 2).
Within the tables the individual weighted sub-scales are reported, as well as the total weighted work-load. The perceived workload was \textit{`significantly lower'} (paired t-test, p=0.001) in the proposed tool than the clinical tools perceived by both experts. It is worth noting a lower score for perceived performance in this table corresponds to a better perceived performance. Therefore, the experts found the proposed tool to perform better than the clinical tools and UGIR. Furthermore, the perceived mental workload of the proposed tool improved at the end of the experiment, showing improved performance after exposure to the tool.  
%\begin{minipage}
%\begin{minipage}[t]{1\linewidth}
\begin{table}[t]
\centering
\caption{The perceived weighted workload score from the NASA-TLX questionnaire and individual sub-scale scores that contribute to it are shown. A low weighted workload score corresponds to a tool that requires less effort, less frustration, less mental, temporal and physical demands, and a tool that has a higher perceived performance.}

\begin{tabular}{c|cccc}
\hline
NASA-TLX        & \multicolumn{4}{c}{Average}   \\ \cline{2-5} 
weighted scores & Trace & Point & UGIR  & Beyond \\ \hline
Effort          & 13.67 & 15.50 & 14.00 & 6.17   \\
Frustration     & 12.00 & 11.00 & 21.34 & 2.17   \\
Mental Demand   & 14.33 & 10.84 & 8.67  & 6.17   \\
Performance     & 11.00 & 18.34 & 28.34 & 5.34   \\
Physical Demand & 6.67  & 4.67  & 4.00  & 3.67   \\
Temporal Demand & 0.84  & 0.00  & 0.00  & 0.34   \\ \hline
Total workload  & 58.47 & 60.34 & 76.33 & 23.84  \\ \hline
\end{tabular}
\end{table}
%\end{minipage}
\section{Discussion} 
Fig. \ref{fig:images} shows visually similar contours for all tools. The proposed method shows more anatomically plausible results than the Point tool. 
%The impact BEAS has on the CNN segmentation can be seen in fig. \ref{fig:images}, where the boundary is more anatomically plausible than the point tool. 
%The image quality is slightly reduced in Beyond, as the 2D plane viewed in 4D View software has been converted into a nifti format. 4D View also allows the user to view the 2D US image and corresponding coronal and mid-saggital planes. This is useful in segmentation as the structures and features that define the levator hiatus boundary can be viewed in the corresponding planes if noise or artefacts are present. Therefore, to achieve a good result on Beyond from a 2D US image alone may be assumed to be more difficult. Therefore, the results may improve further if this tool was to be implemented on 4DView within the clinical setting. 
The proposed tool achieved visually `clinically acceptable' results for almost all cases, however, only 2 and 1 CNN segmentations required no editing from expert 1 and 2 respectively. The sub-optimal segmentation was due to a poor US acquisition (that was noted as a clinically unacceptable US image), this reduced the visibility of the levator hiatal boundary, which severely impacted the CNN output. Retrospectively, the radius defining the initial BEAS contour in the second task was increased, and an optimal segmentation was obtained using the proposed tool.  Thus, the tool is capable of a `clinical acceptability' score of 100$\%$, with further tuning. 

Both experts achieved at least an average improvement of 13 seconds when using the proposed tool. The time measured did not include CNN inference time, to keep it independent to the experiment, and to allow for direct comparison with other segmentation tasks of different CNN architectures. It is assumed with optimisation the CNN inference time would reduce.  The proposed tool nonetheless, is still quicker than the clinical tools and UGIR, and it may be assumed with further practice, the time taken would continue to decrease. 
%The proposed tool also reduces subjective factors, such as; frustration, mental and temporal demand.  
%Regarding the effect of the proposed tool on the perceived mental workload, the results from Table 1, show an observed decreased mental workload in the segmentation of the levator hiatus compared to current clinically used tools.

Following 30 levator hiatus segmentations, the NASA-TLX questionnaire demonstrated the tool improved perceived performance, reduced effort, frustration, mental and temporal demand. The proposed tool reduced the weighted perceived workload, by 36.50, 34.63 and 52.49 points on the NASA-TLX index scale, for Point, Trace and UGIR tools respectively. The performance improved by 13.00, 5.66, 23.00 points on the NASA-TLX weighted index scale compared to the Point, Trace and UGIR tools respectively. It may be assumed the performance of the Point tool was lower, due to the less anatomically accurate segmentation, shown in Fig. \ref{fig:images}. Therefore, it may be assumed that this work could improve the clinical workflow. 
%The perceived weighted workload score of the proposed tool was substantially lower than the clinical tools. 
Table 2 showed that the perceived workload score was lower at the end of the experiment than at mid-experiment, highlighting that with increased exposure, the workload may continue to reduce. 
\begin{table}[t]
\centering
\caption{The perceived weighted workload score from the NASA-TLX questionnaire and individual sub-scale scores that contribute to it are shown from mid-experiment (attempt 1) and the end of the experiment (attempt 2). A low weighted workload score corresponds to a tool that requires less effort, less frustration, less mental, temporal and physical demands, and a tool that has a higher perceived performance.}

\begin{tabular}{c|ccc|ccc}
\hline

 {NASA-TLX} & \multicolumn{3}{c|}{Attempt 1} & \multicolumn{3}{c}{Attempt 2} \\ \cline{2-7} 
weighted scores                                         & Trace    & Point    & Beyond   & Trace    & Point    & Beyond   \\ \hline
Effort                                                  & 11.00    & 14.67    & 7.33     & 16.33    & 16.33    & 5.00     \\
Frustration                                             & 12.00    & 11.00    & 3.33     & 12.00    & 11.00    & 1.00     \\
Mental Demand                                           & 15.33    & 9.67     & 5.33     & 13.33    & 12.00    & 7.00     \\
Performance                                             & 10.67    & 15.00    & 6.00     & 11.33    & 21.67    & 4.67     \\
Physical Demand                                         & 7.00     & 4.00     & 3.33     & 6.33     & 5.33     & 4.00     \\
Temporal Demand                                         & 1.67     & 0.00     & 0.00     & 0.00     & 0.00     & 0.67     \\ \hline
Total workload                                          & 57.60    & 54.34    & 25.34    & 59.33    & 66.33    & 22.33    \\ \hline
\end{tabular}
\end{table}

The proposed tool is compounded of a post-processing filter and an interactive algorithm. It can be easily implemented on other 2D segmentation tasks, to improve the segmentation boundary and allow for easy editing of incorrect segmentation. There is scope to expand this work to 3D segmentation. Currently, the hyper-parameters used for BEAS (i.e. number of B-splines) requires manual optimisation. In future work, it would be beneficial to automate hyper-parameter selection dependent on the initial 2D segmentation, and compare performance for several segmentation tasks.% In addition, 
%In the future it would be, this work is only implemented in 2D, it should be possible to adapt Beyond to handle 3D segmentation editing, albeit it will not be trivial for GUI adaptation. 
%However, the BEAS post-processing in 3D would be more trivial to implement, as BEAS has been commonly used in 3D segmentation tasks in literature \cite{beasbarboa2012,pedrosabeas2017,7586276}. In future work, it may be possible to use clinical adaptations of 2D segmentations in beyond to better train the initial CNN, i.e. by incorporating a heat-map that focuses on these regions in training.
\section{Conclusion}
To conclude, in this work, a novel CNN-based interactive 2D segmentation tool was proposed. The interactive element works in real-time and requires less user time and perceived workload than current clinical methods, suggesting the proposed work may improve the current clinical workflow. The method utilised the BEAS framework, which ensured the final contour was more biologically plausible than CNN segmentation outputs. This framework can easily be implemented for other 2D segmentation tasks, to make the results more robust while improving the clinical acceptability and giving liability to clinicians.   %The proposed method scored a lower perceived workload score than the current clinical methods, suggesting it may improve the current clinical workflow.   

\section{Acknowledgments}
We gratefully acknowledge General Electric Healthcare (Zif, Austria) , for their continued research support. 
%
% the environments 'definition', 'lemma', 'proposition', 'corollary',
% 'remark', and 'example' are defined in the LLNCS documentclass as well.
%%
%F
\bibliographystyle{abbrv}
\bibliography{paper739}
%d \begin{thebibliography}{8}
%Author, F.: Article title. Journal \textbf{2}(5), 99--110 (2016)

%Author, F., Author, S.: Title of a proceedings paper. In: Editor,
%F., Editor, S. (eds.) CONFERENCE 2016, LNCS, vol. 9999, pp. 1--13.
%Springer, Heidelberg (2016). \doi{10.10007/1234567890}

%Author, F., Author, S., Author, T.: Book title. 2nd edn. Publisher,

%Author, A.-B.: Contribution title. In: 9th International Proceedings
%on Proceedings, pp. 1--2. Publisher, Location (2010)

%LNCS Homepage, \url{http://www.springer.com/lncs}. Last accessed 4
%\end{thebibliography}
\end{document}